 \documentclass[preprint2]{emulateapj}

  \usepackage{amsmath,color,natbib}

  \renewcommand{\d}{\mathrm{d}}       

  \newcommand{\vct}[1] {\boldsymbol{#1}}
  \newcommand{\uv} [1] {\hat{\vct{#1}}}        

\slugcomment{\today, Ver 0.8a}
\begin{document}


\title{Local anisotropy, higher order statistics, and turbulence
  spectra}


\author{W. H. Matthaeus\altaffilmark{1},
        S.  Servidio\altaffilmark{2},
        P.  Dmitruk\altaffilmark{3},
        V.  Carbone\altaffilmark{2},
        S.  Oughton\altaffilmark{4},
        M.  Wan\altaffilmark{1},
and
        K.  T.  Osman\altaffilmark{1}}

\altaffiltext{1}{Department of Physics and Astronomy,
        University of Delaware, Delaware, USA}
\altaffiltext{2}{Dipartimento di Fisica, 
Universita della Calabria, I-87036 Cosenza, Italy}
\altaffiltext{3}{Departamento de F\'\i sica
        (FCEN-UBA \& IFIBA-CONICET),
        Buenos Aires, Argentina}
\altaffiltext{4}{Department of Mathematics,
                  University of Waikato,
                  Hamilton 3240, New Zealand}

\begin{abstract}
 Correlation anisotropy emerges dynamically in magnetohydrodynamics
(MHD),
producing stronger gradients across the large-scale mean magnetic field than
along it.  This occurs both globally and locally, and has significant
implications in space and astrophysical plasmas, including
particle scattering and transport, and theories of turbulence.
Properties of local correlation anisotropy
are further documented here by showing through numerical experiments that
the effect is intensified in  more localized estimates of the mean field.
The mathematical formulation of this property shows that
local anisotropy mixes second-order
with higher order correlations.
Sensitivity of local statistical
estimates to higher order correlations
can be understood in connection with
the stochastic
coordinate system inherent in such formulations.
We demonstrate this in specific cases,
illustrate the connection 
to
higher order statistics 
by showing the sensitivity of local anisotropy
to phase randomization, and thus
establish
that 
the local structure function 
is not a measure of
the energy spectrum.
Evidently the local enhancement of correlation
anisotropy is of substantial fundamental interest,
and this phenomenon must be understood in terms of higher order correlations,
fourth-order and above.

\end{abstract}


\maketitle

\section{Introduction: correlations and anisotropy}
        \label{sec:intro}

Correlation and spectral anisotropy play
important roles in
current studies of solar wind
and astrophysical plasmas, having
significant impact on
descriptions of
the turbulence cascade, particle scattering,
the nature of kinetic dissipation, and the
transport of turbulence.
Evidence from experiments, numerical simulations,
theoretical arguments, and spacecraft observations
has consistently supported the conclusion that
magnetohydrodynamic (MHD)
turbulence leads to states characterized by
gradients that are relatively
stronger when
measured perpendicular
to the large-scale magnetic field
and relatively
weaker when measured parallel to it
  \citep{RobinsonRusbridge71,ZwebenEA79,RenEA11,
        ShebalinEA83,Bondeson85,CarboneVeltri90,OughtonEA94,
        MontTurner81,MattEA90,BieberEA96}.
This
        \emph{correlation anisotropy}
has been quantified both globally and locally,
by varying the definition of the
mean magnetic field.
The local form, being of greater magnitude,
has received substantial attention in recent
years
        \citep{ChoVishniac00-aniso,MilanoEA01,
                HorburyEA08,TesseinEA09,Podesta09-SWaniso,LuoWu10,%
                WicksEA10-slopes,WicksEA11,ChenEA11-SWaniso}.
The present paper will focus
on the
relationship between these forms
of anisotropy,
providing a better characterization of
the scale dependence of local anisotropy.
A particular conclusion will be
that correlation anisotropy
affects statistics at all orders,
including but not limited to the
energy spectrum and other second-order
statistics.
We also present evidence that the enhancement of
local anisotropy over the global value is
due entirely to higher order statistical effects.

The basic physics leading to this strong perpendicular spectral transfer
and enhancement of perpendicular anisotropy
has been elucidated in the context of
incompressible
homogeneous MHD turbulence
        \citep{MontTurner81,ShebalinEA83,Bondeson85,Grappin86,OughtonEA94}.
For this model, all modal couplings are
triadic
        (involving wavevectors
        $\vct{k}_1$,  $\vct{k}_2 $,  $\vct{k}_3 $,
such that
        $\vct{k}_1 + \vct{k}_2 + \vct{k}_3 = 0 $),
but in the
presence of a strong uniform DC magnetic field
        $ B_0 \uv{z} $,
only those couplings that have
one
        (or all three)
wavevectors perpendicular to $ \uv{z}$
will proceed unattenuated by propagation effects.
All other couplings are suppressed
to a progressively greater degree as $B_0$ is strengthened.
The appearance of stronger perpendicular
gradients
implies that
spectral transfer to the perpendicular
wavevectors is more robust than that to the parallel
wavenumbers.
This corresponds to appearance of
characteristic
        \emph{spectral anisotropy},
with enhancements of high  $ k_\perp $
power relative to
        $k_\parallel $
power,
an effect
that become progressively stronger at smaller scales
        \citep{ShebalinEA83,OughtonEA94}.
This organization of the spectrum, is built into
models such as Reduced MHD (RMHD)
        \citep{KadomtsevPogutse74,Strauss76,Mont82-strauss}
and related models
such as
the
        \cite{GoldreichSridhar95}
steady-state (GS) model,
and the nearly incompressible MHD (NIMHD) model
        \citep{ZankMatt92b}.
Indeed strong anisotropy of MHD correlations
is observed in laboratory devices
        \citep{RobinsonRusbridge71,ZwebenEA79,RenEA11},
in the solar wind
        \citep{MattEA90,TuMarsch95,BieberEA96}
and the corona
        \citep{ArmstrongEA90},
and is an apparent requirement for
consistency of scattering theory with solar energetic
particle (SEP) observations
        \citep{BieberEA94,Droege05}.

However anisotropy of the energy spectrum is not the only effect
associated with correlation anisotropy and
enhancement of perpendicular gradients.\footnote{
Here we are concerned with correlation
(or spectral) anisotropy, e.g., anisotropy 
of $|{\bf b}({\bf k})|$ for varying ${\bf k}$.
Anisotropy of the {\it variance}, i.e., direction of 
${\bf b}({\bf x})$, also occurs, 
see \citet{BelcherDavis71}, but is a distinct issue,
which however may become linked in certain theories, e.g., 
\cite{ZankMatt92b,GoldreichSridhar95}.}
For example, one can discuss
the shape and
orientation of    \emph{structures},
such as ``eddies''
        \citep{ChoVishniac00-aniso}
or current sheets
        \citep{DmitrukEA04-testpart}.
It is well known that one of the
powerful nonlinear effects of turbulence is the production
of coherent structures that are progressively smaller
at small scales.
This leads to intermittency as measured
by nonGaussianity and higher order statistics,
as seen in hydrodynamic and MHD models
        \citep{SheLeveque94,PolitanoPouquet95}.
Simulations show that such nonGaussian statistics are generated
very rapidly by the cascade of excitations to smaller scales
        \citep{WanEA09-nongauss}.
In MHD,
under a wide variety of conditions,
the associated coherent structures take the form of enhancements of
electric current density in sheets or filaments
        \citep{MattLamkin86,CarboneEA90,Veltri99}.
When the turbulence is anisotropic relative to a large-scale
field
        \citep[e.g.,][]{DmitrukEA04-testpart},
the
current structures tend to align in that direction as well.
This suggests that higher order statistical quantities
(fourth-order moments, kurtosis, etc) also must become anisotropic.

The above reasoning leads
to an expectation that statistics
at all orders might
be involved in correlation anisotropy,
but many
        ``theories of MHD turbulence''
discuss exclusively the
properties of the second-order statistics; that is,
the energy spectra and associated correlation and structure functions.
This emphasis follows naturally from
the use of
wavevector spectra and two-point, single
time correlation functions
as
measures of the
distribution of energy in spatial structures of
varying size
        \citep{MoninYaglom},
as exemplified by the classic
        \cite{Kol41a}
theory.
Recognizing this background, it is not
entirely surprising that recent interest in local
anisotropy has sometimes focused on
an interpretation of this effect as a
        \emph{local energy spectrum}
        \citep{ChoVishniac00-aniso,ChenEA11-SWaniso,WicksEA11}.
The discussion below provides
in effect a critique of this
interpretation.

To render the discussion specific,
we assume a statistical description
that is
homogeneous in space and stationary in time.
The brackets
        $ \langle \cdots \rangle $
denote an ensemble average,
which by invoking a classical ergodic theorem
is approximated in practice by a space or time average.
It may be possible to define
the statistical ensemble in other ways,
but here
a classical
statistical framework
is assumed
        \citep{MoninYaglom}.
The two-point correlation
function measuring the statistical relationship
between the magnetic field fluctuation at points $\vct{x}$
and displaced position $\vct{x}^\prime = \vct{x} + \vct{r}$
is also the Fourier transform of the
wavevector energy spectrum,\footnote{%
    For three dimensional
    isotropic Kolmogorov turbulence the omnidirectional energy
    spectrum is
        $ {\cal E}(k) = 4 \pi k^2 S_{\alpha \alpha}(\vct{k}) $
     (sum on repeated indices),
    with
        $ {\cal E}(k) \sim k^{-5/3}$
    in the inertial range.}
 $ S_{\alpha\beta}(\vct{k}) $.
The definitions are
        \citep[e.g.,][]{MoninYaglom}
\begin{equation}
     R_{\alpha \beta}(\vct{r})
    =
     \langle b_\alpha b'_\beta  \rangle
  =
     \int \d^3 k \,
        \ S_{\alpha \beta} (\vct{k}) e^{i \vct{k} \cdot \vct{r}} ,
\label{eq:S}
\end{equation}
where we abbreviate
        $ b_\alpha =  b_\alpha (\vct{x}, t) $
and
        $ b'_\beta =  b_\beta (\vct{x} + \vct{r}, t )$,
and suppress the time argument.\footnote{%
   In the solar wind
        \citep{Jokipii73},
   and wind tunnels
        \citep{MoninYaglom},
   rapid sweeping
   past detectors provides a useful connection between
   temporal and spatial statistics, through the Taylor frozen-in flow
   hypothesis.  Here we will not be concerned with time
   statistics.}
Another two-point statistic is the second-order
structure function
\begin{equation}
   D(\vct{r})
      =
   \langle  | \vct{b} - \vct{b}' |^2
   \rangle
     \equiv
    2 \delta b^2 - 2 R( \vct{r} ) ,
 \label{eq:d}
\end{equation}
which is obviously related to $R_{\alpha\beta} $.
Here
        $ \delta b^2 = R_{\alpha\alpha}(0) $
is the total variance
of the fluctuations
and
        $ R(\vct{r}) = R_{\alpha \alpha}(\vct{r})$.

Existence of a homogeneous statistical
ensemble implies
that there exists a probability density ${\cal P}(\Gamma)$
that describes
fully all realizations of the turbulence, where
each realization is labeled by an index
$\Gamma$.
This density may
be projected onto the
space of two-point statistics,
arriving at a probability distribution function
(pdf)
        $ P_2 (\vct{b}, \vct{b}') $
that characterizes
all of the two-point statistics of any order.
Thus one may equivalently express the structure function
in two ways as
\begin{eqnarray}
  D(\vct{r})
        & = &
  \int_\Gamma \d\Gamma \,  \left [|\vct{b} - \vct{b}^\prime |^2
                        \right ]_\Gamma  {\cal P}(\Gamma)
  \\
        & = &
  \iint \d^3 b \,  \d^3 b' \,
  \left| \vct{b} - \vct{b}' \right|^2
          P_2( \vct{b}, \vct{b}'),
 \label{eq:P}
\end{eqnarray}
where
        $\left [ |\vct{b} - \vct{b}'|^2 \right ]_\Gamma$
is the
value of the square-bracketed quantity for realization $\Gamma$.
Similar relations hold for the spectra and two-point
correlation functions.
Note that the structure function $D$
and the correlation matrix $R_{\alpha \beta}$
are
        ``second-order statistics''
because they are explicitly written as
second-order moments of either the full density
        $\cal P$
or of the reduced two-point probability density         $ P_2 $.

In general the turbulence may not be isotropic, especially if
        $ \langle \vct{B} \rangle = \vct{B}_0 \ne \vct{0} $.
It is then of interest to compute correlations
in directions relative to
        $ \uv{B}_0 = \vct{B}_0 / B_0 $
(and a perpendicular direction  $ \uv{n}_0$).
We will also
consider a
        \emph{locally defined}
mean field
in which the local mean direction
        $ \uv{H}
        = \vct{H}/ |\vct{H}| $
is itself a
random variable that satisfies
        $ \langle \vct{H} \rangle= \vct{B}_0$.
We have in mind several possible ways to determine
        $ \vct{H}$,
for example as
        $ \vct{H}_V(\vct{x}) = V^{-1} \int_V \d^3r \, \vct{B}$,
defined in subvolumes or ``boxes'' of
volume
        $ V $  
centered on $\vct{x} $;
or as
        $ \vct{H}_L = L^{-1} \int_L \d r \, \vct{B}$,
defined in co-linear samples of length $L$
  (as used in
   single spacecraft estimates in the solar wind,
   under Taylor's hypothesis);
or
perhaps as
        $ \vct{H}_{\vct {r}} (\vct{x}) $
defined as the average of $\vct B$
at ${\vct x}$ and $\vct {x}'$,
$\frac{1}{2}
                   \left[
                \vct{B}(\vct{x} + \vct{r}) + \vct{B}(\vct{x})
                   \right] $.
These may be called, respectively, volume averages,
line averages, and point averages.

At this point we recall
a generalization of standard
structure functions
        \citep{MilanoEA01}
that formally coincides with $D(\vct{r})$,
\begin{equation}
   \tilde D (s)
        \equiv
    \langle |
        \vct{b}( \vct{x} ) - \vct{b}( \vct{x} + s \uv{c} )
            |^2 \rangle,
\label{eq:tS}
\end{equation}
except that the direction $\uv{c}$ is permitted
to be a variable direction, in particular a direction determined
relative to the local estimates $\uv{H}$.
For special cases,
the separation
can be chosen
to be either parallel to  the local mean
magnetic field,
        $ \uv{c} \to  \uv{H} $,
or else
perpendicular to        $ \vct{H} $:
        $\uv{c} \to \uv{n}
           \equiv
        \uv{e} \times \uv{H}  / | \uv{e} \times \uv{H} | $
for reference direction $\uv{e} $.
This enables us to define the
(locally) parallel structure function
$\tilde D^\parallel(s)
        =
  \langle |\vct{b} (\vct{x}) - \vct{b} (\vct{x} + s \uv{H})|^2 \rangle $
and the local perpendicular (transverse) structure function
  $ \tilde D^\perp(s)
        =
    \langle | \vct{b }(\vct{x}) - \vct{b}(\vct{x} + s \uv{n})
            |^2
    \rangle $.
Our main concern here will be to
better understand the
relationship between local and global anisotropy, and
between
        $D$ and $\tilde D$.
We conclude that
however appealing their similarity might be,
one can anticipate
fundamental differences.
Notably, since
        $ \tilde D $
is defined in a
        {stochastic coordinate system}
it is thus a \emph{conditional} statistic.
      $ \tilde D$
is therefore necessarily a higher order statistical quantity.
In contrast,
        $ D $
is defined
in a fixed coordinate system and is
related directly to the
second-order statistics, such as
correlation functions and spectra.

  \citet{MilanoEA01}
showed that structure functions
defined as above can be employed
to quantify both global anisotropy
and local anisotropy.
For separations $s$
lying in the inertial range and smaller,
and for cases with nonzero $B_0$,
global anisotropy takes the
form
        $ D( s \uv{n}_0) / D( s \uv{B}_0) > 1$,
which is equivalent to enhanced
perpendicular anisotropy seen in spectra
        \citep{ShebalinEA83,OughtonEA94}.
Local anisotropy
is typically characterized by
        $\tilde D^\perp(s) / \tilde D^\parallel(s) > 1 $,
which corresponds to stronger gradients
perpendicular to a local preferred direction.
Interestingly, the local anisotropy was found to be greater
than the global anisotropy at a given scale,
    $ \tilde D^\perp(s) /  \tilde D^\parallel(s)
        >
             D( s \uv{n}_0 ) / D( s \uv{B}_0 )
        >  1
    $
 \citep{MilanoEA01}.
It is noteworthy that local anisotropy
was found
even in cases that are globally isotropic
with $B_0 = 0$.
A related study
        \citep{ChoVishniac00-aniso}
found similar local anisotropy, and
argued for
a correspondence to anisotropic magnetic spectra
in the
        ``reduced MHD'' regime
        \citep{GoldreichSridhar95}.

\section{Results}
   \citet{MilanoEA01} also suggested
that there was evidence
for progressively greater anisotropy as the
mean field is computed more locally.  However the available bandwidth
of the simulations limited the strength of that conclusion.
We present a new result here that demonstrates  this effect
with higher resolution simulations,
employing
        (dealiased)
incompressible 3D MHD spectral method simulations at
a spatial resolution of         $ 512^3 $.
Thus the maximum retained wavenumber is
        $ k_{\text{max}} = 170 $
        ($ k = 1$ corresponds to
         the longest allowed wavelength
         in the periodic box).
The runs are initial value problems, decaying turbulence with initially
excited wavenumbers of
        $k = 1$ to $6$,
and
initial fluctuation energy normalized such that
        $ \langle \vct{b} \cdot \vct{b} \rangle =
          \langle \vct{v} \cdot \vct{v} \rangle
                = 1 $.
The resistivity     $ \mu$
and viscosity   $ \nu$,
are selected to ensure good spatial resolution,
meaning that
        $ k_{\text{max}} / k_{\text{diss}} > 3 $
at all times
        \citep{WanEA10-accuracy}.
The dissipation wavenumber is defined as
        $  k_{\text{diss}}
         = 1 / \lambda_{\text{diss}}
         = (\epsilon/ \nu^3)^{1/4} $,
where
        $ \epsilon $
is the rate of (total) energy dissipation.
Initial Reynolds numbers are
        $ R = 1 / \nu   =  Rm = 1 / \mu = 2000 $.

We report on two runs, one with
        $B_0 = 0 $
and another with $B_0 = 1$.
Broadband energy spectra develop and the data are analyzed
near the time of maximum dissipation.
We examined
the anisotropy relative to
local
mean fields determined
variously as volume, line and point averages.
Corresponding
estimates of
        $ \tilde D^\parallel (s)$
and
        $ \tilde D^\perp (s)$
were computed in the same way.
For volume averages, or line averages
with         $ V^{1/3} \sim L  \gg \lambda_c$,
the correlation scale of the fluctuations,
we expect that
$ \vct{H}_V$ provides increasingly
refined estimates of
$\vct{B}_0$.
Indeed, for one dimensional
random process,
the
classical ergodic theorem
        \citep{MoninYaglom}
guarantees that
        $ {\lim}_{L \to \infty}\vct{H}_L
                =
            \langle \vct{B} \rangle  = \vct{B}_0
        $.\footnote{%
     For line averages,
this statistical limit
     is guaranteed
        when the traced correlation function tends to zero
        at large lags
                $ |\vct{r}| \to \infty $
        fast enough so that
                $ \int_0^\infty \d x \, x R(x,0,0) $
        is bounded for any choice
        of Cartesian axis $x$.}

To assess the degree of local anisotropy, the
local structure functions are computed
a locally-determined
mean-field-aligned axisymmetric coordinate system.
That is, estimates of
        $ \tilde D^\parallel(s)$
and     $ \tilde D^\perp(s)$
are accumulated and averaged.
Several methods are used to determine the local
mean field -- volume, line and point averaging, as described above.
Figure~1 shows the ratio
$  \tilde D^\perp(s) /\tilde D^\parallel(s)$
which provides a measure of the
variation of anisotropy with lag $s$,
for several different
determinations of
the local mean field.
Also shown is the anisotropy ratio
relative to the global
mean magnetic field.
It is clear that
(1)
the degree of anisotropy
is greater at smaller lags;
and (2) that anisotropy is
more pronounced
when
the mean magnetic field is computed over a smaller region.
This was found by
        \citet{MilanoEA01} and
        \citet{ChoVishniac00-aniso}
and confirmed in numerous other
studies.
As has been reported previously,
local anisotropy is present
not only for the nonzero $B_0$ case (Fig. 1b)
but also
for zero global mean field (Fig 1a).
We thus
confirm that
stronger gradients are produced perpendicular to the magnetic field,
and also that local anisotropy is stronger than global anisotropy.

Clearly the local structure functions are of great interest,
and we now discuss their formal nature.
It is convenient to introduce the
generalization
  $ \tilde D (s_\parallel,s_\perp)
        =
    \langle
        | \vct{b}(\vct{x}) - \vct{b}(\vct{x} + s_\parallel \uv{H}
          + s_\perp \uv{n})|^2
    \rangle $,
which makes explicit the
possibility of arbitrary separations, both
parallel ($s_\parallel$) to and perpendicular ($s_\perp$) to
the local
mean magnetic field.
Evidently
        $ \tilde D^\parallel(s) = \tilde D(s,0) $
and     $ \tilde D^\perp(s) = \tilde D(0,s) $.

The quantity $\tilde D$
can be
        \emph{formally}
obtained from the expression for
$D(\vct{r})$ in Eq.~(\ref{eq:d})
by
the replacement
        $ \vct{r} \to s_\parallel \uv{H} + s_\perp \uv{n} $
in Eq.~(\ref{eq:P}).
However some care is required in interpreting
this procedure due to the new stochastic variables that appear
in the arguments.
(For example, the implication is that
the differential        $ \d^3 b^\prime$
becomes stochastic.)
To avoid complications we choose to write
the defining relation in terms of the
full probability density of realizations,
that is,
\begin{equation}
   \tilde D(s_\parallel,s_\perp)
 =
   \int_\Gamma
     \d \Gamma
      \left [ |\vct{b}(\vct{x} + s_\parallel \uv{H}
          + s_\perp \uv{n}) - \vct{b}(\vct{x})|^2
      \right ]_\Gamma
     {\cal P}(\Gamma).
 \label{eq:tildeDP}
\end{equation}
This new expression is not a simple coordinate transformation of
Eq.~(\ref{eq:P}) because the coordinate unit vectors
        $ \uv{H} $ and
        $ \uv{n} $
are themselves random variables, dependent upon $\Gamma $.
However the new
independent variables $s_\parallel$ and $s_\perp$ are non-random
coordinates in the stochastically rotated reference frame of the local
systems.  It is clear that
        $ \tilde D (s_\parallel, s_\perp) $
remains a
second-order statistical quantity only if the coordinate system is
fixed and
        $ \uv{H} $ and $ \uv{n} $
become fixed  vectors, or
if the probability distributions are insensitive to directions.  The
former case is obtained asymptotically as $B_0 \to \infty$
and the
mean field direction becomes statistically sharp.  The latter case is
obtained for highly symmetric cases such as isotropic turbulence.  In
general, unless the density $\cal P$ is invariant under the stochastic
changes of $\vct{H}$, the statistical nature of $\tilde D$ becomes
higher order.

\begin{figure}
 \begin{center}
  \includegraphics[width=0.8\columnwidth]{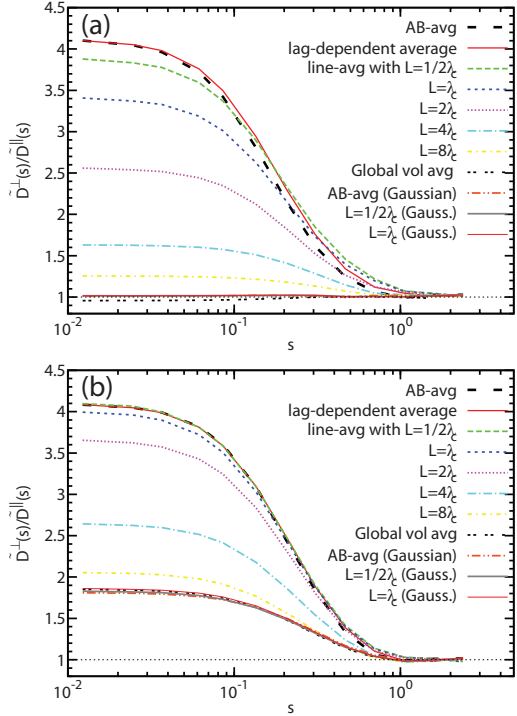}
 \end{center}
\caption{Anisotropy ratio
        $ \tilde D^{\perp}(s) / D^{\parallel} (s) $
    vs. lag $s$ at inertial range and larger scales.
    Point and line average
estimates of local mean field $ \vct{H} $ are used
(``AB'' averages the two points $\vct{x}$, $\vct{x}'$,
``lag dependent'' integrates from $\vct{x}$ to $\vct{x}'$.)
Perpendicular excitation is relatively enhanced, except for lags
        $ s \gtrsim 1$
    where isotropy holds (N.B. energy containing
eddies are initially isotropic).
    (a) Results from $B_0 = 0$ simulation, with
                $ \lambda_c = 0.34 $
     and        $ \lambda_{\text{diss}} = 0.021 $.
     Anisotropy is stronger for more localized estimates
of $\vct{H}$ (e.g., shorter line averages).
Randomizing phases (Gaussianizing --
see text), produces isotropic results for
all averaging methods and at all lags.
Therefore, anisotropy is
associated with non-Gaussian correlations.
     (b) Results from $B_0 = 1$ simulation, with
                $ \lambda_c^\perp = 0.34 $,
                $ \lambda_c^\parallel = 0.43 $,
     and        $ \lambda_{\text{diss}} = 0.022 $.
This case is anisotropic relative to the globally determined
$\vct {B}_0$,
and all methods based on
local determinations of the mean field show further
enhancement of anisotropy.
For this run, phase randomization removes the
enhanced local anisotropy, but global anisotropy remains.
  \label{fig-1}
}
\end{figure}

The latter point and its consequences can be made more explicit by
considering special cases.  For small separations purely in the
parallel direction
        ($s_\perp = 0$ and small $s_\parallel$),
and considering the decomposition
        $ \vct{H} \equiv \vct{B}_0 + \vct{h} $,
we may expand
\begin{eqnarray}
  \tilde D(s,0)
         & = &
     \langle
        | s \uv{H} \cdot \nabla \vct{b}( \vct{x} )
         +
          \frac {s^2}{2}
             \uv{H}_i \uv{H}_j \nabla_i \nabla_j \vct{b}( \vct{x} )
         +
           {\cal O}(s^3)
        |^2
     \rangle
        \nonumber \\
        & = &
      s^2 \langle
                  {| \widehat{(\vct{B}_0 + \vct{h})} \cdot \nabla \vct{b} |^2 }
          \rangle
        +
          {\cal O} (s^3) .
 \label{eq:expand}
\end{eqnarray}
Suppose that $|\vct{b}|/B_0  \ll 1$
is small enough to justify an additional
expansion in that parameter.
This yields
        $ \tilde D(s,0)
                =
          s^2 \uv{B}_{0i} \uv{B}_{0j}
          \langle (\nabla_i b_k) (\nabla_j b_k) \rangle
         +
          {\cal O}( h b b / B_0)
        $.
The first term
is a second-order moment.
The remaining terms include third, fourth, and higher-order moments.
Only in the asymptotic limit,
in which
        $ \vct{H} \to \vct{B}_0 $,
is a second-order moment recovered.

As a second special case suppose that
  (i) $B_0 = 0$, so the global anisotropy
may be zero, and
  (ii) that the local mean field estimate is
completely localized so that
        $ \vct{H} = \vct{h} = \vct{b}$.
Then once again
beginning with the small $s$ expansion
Eq.~(\ref{eq:expand}),
we find that
  $ \tilde D(s,0)
       =
     s^2 \langle | \vct{b} \cdot \nabla \vct{b}|^2/|\vct{b}|^2 \rangle
        + {\cal O}(s^3)
  $,
which involves moments higher than second-order.
Even in the very special case that the magnetic field
fluctuation is ``arc polarized''
and has constant magnitude $\sigma$,
this still only reduces to a fourth-order quantity
  $ \tilde D(s,0)
       \approx
    (s^2/\sigma^2)
       \langle | \vct{b} \cdot \nabla \vct{b} |^2 \rangle $.

Observing that the relationship between the
structure function $D(\vct{r})$ and
the local field aligned structure function
$\tilde D$ is somewhat analogous in form to the
relationship between
Eulerian and Lagrangian correlation functions,
one might be tempted to seek similar approximations
to connect them.
One possibility, based
on the ideas underlying Corrsin's
hypothesis of independence
        (see, e.g., \citet{McComb})
is to treat the distribution of the
magnetic field fluctuations as independent of the
distribution of the mean magnetic field directions.
Thus, symbolically one might attempt an
approximation such as
        $P_3 (\vct{b}, \vct{b}', \uv{H})
                \to
         P_2 (\vct{b},\vct{b}') P_H( \uv{H}) $,
from which it would follow that
$\tilde D (s_\parallel, s_\perp)
        =
  \int \d\Omega_H \,
        C ( s_\parallel \uv{H} + s_\perp \uv{n} )
        P_H ( \uv{H} ) $,
where $P_H( \uv{H})$ is the distribution of
mean field directions and
        $ \d \Omega_H$
the differential
of solid angles associated with those directions.
However such an approximation must fail, as can be readily seen:
Suppose that $ C(\vct{r}) $
has a strong perpendicular anisotropy.
Then the random distribution of mean field directions $P_H$
will dull the sharpness of the anisotropy by averaging parallel separations
with perpendicular separations.  This independence hypothesis
would thus produce local anisotropy that is weaker than the global anisotropy.
This is inconsistent with both simulation results and solar wind
results and therefore this approximation is invalid for
the turbulence of interest.

In order
to demonstrate their dependence on higher order correlations,
we examined numerically
the effect on local anisotropy
of a Gaussianization or phase-randomization
process.  This procedure
was carried out for the
same simulation
dataset
described above.
In particular for both the $B_0 = 0$ and
$B_0=1$ cases,
we modified the Fourier coefficients
by randomizing their phases while keeping their magnitudes unchanged.
The effect of phase randomization
is to produce a signal that is Gaussian, lacking coherency associated
with phase correlations.  However this process does
not modify the energy spectrum.
Employing the phase randomized signal, we again compute the
locally defined structure functions using the same
set of methods for determining the local mean field
that was described earlier.
The scaling of the anisotropy with lag is shown
in Fig.~\ref{fig-1}, where it is compared with the original
simulation data for which phase coherency, if present,
was maintained.
The phase randomization has
a dramatic effect in both cases:
when there is global isotropy, the phase randomization
completely destroys the local anisotropy.
When there is global anisotropy the phase randomization
completely eliminates the local
   \emph{enhancement} of anisotropy.

        \section{Discussion}

The issues described above appear to have immediate relevance to a
number of analysis procedures that rely on local accumulation of data.
For example,
even apart from the issue of anisotropy,
the discussions presented here are also relevant to studies that
for various reasons
employ short datasets to define a ``mean''
magnetic field and similarly local correlation or structure
functions.\footnote{%
        This may be motivated by interest in local fluid physics, local
        kinetic physics, or simply lack of available high-cadence data.}
Invariably use of short intervals
        (less than $\lambda_c$, or its temporal equivalent)
time   series
leads to poorly determined global statistics, although an
interpretation in terms of local conditional statistics may still be
meaningful
   \citep[e.g.,][]{SahraouiEA09,SahraouiEA10-subproton,
        AlexandrovaEA11-preprint}.
However, the connection to various orders of statistics
needs to be established.

Another popular approach is to employ wavelet techniques to
determine
        ``local spectra''
as well as measures of local anisotropy analogous to
        $ \tilde D $
        \citep[e.g.,][]{HorburyEA08,WicksEA10-slopes,WicksEA11,ChenEA11-SWaniso}.
This approach is also based on conditional statistics, since it
determines a local mean field at each scale and then accumulates data
relative to the direction of that local mean field.
It follows that such wavelet spectra will also be higher (than second)
order statistical quantities, and thus are distinct from the actual spectra.

In conclusion,
an examination of structure functions
computed relative to
locally determined mean magnetic field directions
reveals that such quantities
involve higher than second-order moments of the underlying
probability distributions.
This property emerges
because the local mean field direction becomes a random
variable.  Consequently these
structure functions are computed in a stochastic coordinate system,
and involve averaging over the magnetic field at two positions,
        $ \vct{b} $, and    $ \vct{b}' $,
and also the mean magnetic field
direction $ \uv{H} $.
Observing that the energy spectrum and related structure function
are second-order moments, the fact that the locally-oriented
structure functions involve higher order moments implies that
the information they contain is not identical to that of
the energy spectrum.  The additional random degree of
freedom does not appear to be amenable to an adaptation of
the Corrsin independence hypothesis, given that the
local anisotropy is stronger than the global anisotropy.
Gaussianization (phase randomization) of a turbulent field
eliminates the enhancement of the local anisotropy,
confirming its sensitivity not just to higher order correlations,
but in particular to higher order nonGaussian correlations.
Evidently the phenomenon of locally enhanced anisotropy
involves fundamental physics that is embedded in the
higher order statistics, as is the case for intermittency and
coherent structures that are generated by turbulence.
This information is beyond the scope of what can properly be described
with just the energy spectrum.

We thank Randy Jokipii for useful discussions.
This research is partially
supported by
NASA (Heliophysics Theory) NNX11AJ44G, (Heliophysics GI)
NNX09AG31G
and NSF- (Solar Terrestrial) AGS-1063439 \& (SHINE) ATM-0752135,
and by the EU under the 
Marie Curie ``TURBOPLASMA'' project.

  \bibliographystyle{apj}

 \newcommand{\BIBand} {and} 
  \newcommand{\boldVol}[1] {\textbf{#1}} 
  \newcommand{\SortNoop}[1] {} 
  \newcommand{\sortnoop}[1] {} 
  \newcommand{\stereo} {\emph{{S}{T}{E}{R}{E}{O}}} 
  \newcommand{\au} {{A}{U}\ } 
  \newcommand{\AU} {{A}{U}\ } 
  \newcommand{\MHD} {{M}{H}{D}\ } 
  \newcommand{\mhd} {{M}{H}{D}\ } 
  \newcommand{\RMHD} {{R}{M}{H}{D}\ } 
  \newcommand{\rmhd} {{R}{M}{H}{D}\ } 
  \newcommand{\wkb} {{W}{K}{B}\ } 
  \newcommand{\alfven} {{A}lfv{\'e}n\ } 
  \newcommand{\alfvenic} {{A}lfv{\'e}nic\ } 
  \newcommand{\Alfven} {{A}lfv{\'e}n\ } 
  \newcommand{\Alfvenic} {{A}lfv{\'e}nic\ }

\end{document}